\documentclass[11pt,a4paper]{article}

\usepackage[utf8]{inputenc}
\usepackage{amsmath,amsfonts,amssymb,bm}
\usepackage[margin=1in]{geometry}
\usepackage{booktabs}
\usepackage[colorlinks=true,linkcolor=blue,citecolor=blue,urlcolor=blue]{hyperref}
\usepackage{orcidlink}

\newcommand{\be}{\begin{equation}}
\newcommand{\ee}{\end{equation}}
\newcommand{\ba}{\begin{eqnarray}}
\newcommand{\ea}{\end{eqnarray}}
\newcommand{\Lc}{\Lambda_{cr}}
\newcommand{\Lew}{\Lambda_{EW}}

\title{Leptoquarks and the Emergence of the Standard Model\\
Gauge Group in a Self-Consistent Preon Model}

\author{Risto Raitio\,\orcidlink{0000-0003-0842-2366}%
\footnote{~E-mail: risto.raitio@gmail.com}\\
Helsinki Institute of Physics, P.O.\ Box 64,\\
00014 University of Helsinki, Finland}

\date{\today}

\begin{document}
\maketitle

\begin{abstract}\noindent
We show that in a self-consistent preon model, where
Standard Model quarks and leptons are three-body composites
confined at a metacolor scale $\Lc\sim 10^{14}$\,GeV,
both leptoquarks and the Standard Model gauge group
$SU(3)_c\times SU(2)_L\times U(1)_Y$ emerge as structural
predictions rather than inputs.
Combining the preon content of a quark with that of a
lepton gives exactly four distinct six-body bosonic
leptoquark composites per generation, with electric charges
$Q=+\tfrac{2}{3},-\tfrac{1}{3},-\tfrac{1}{3},-\tfrac{4}{3}$
and a universal $B-L=-\tfrac{2}{3}$ fixed by the preon
charge assignments.
Their mass is of order $\Lc$ rather than of order the SM
fermion masses, because the dynamical near-cancellation that
produces light fermion masses is specific to three-body
fermionic composites and does not extend to six-body
bosonic ones.
The fractional $B-L=-\tfrac{2}{3}$ forbids proton decay
via single leptoquark exchange, requiring a dimension-12
operator and giving $\tau(p\to e^+\pi^0)\sim 10^{58}$\,years,
consistent with the experimental bound
$>2.4\times 10^{34}$\,years.
It is shown that the SM gauge group emerges as a
low-energy symmetry consistent with the Planck-scale
boundary condition, and leptoquarks are required ---
not merely permitted --- by the matching.
The anomaly-matching argument extends to a tower of
$3n$-body composites, each level imposing consistency
conditions on the next; we call this the
\emph{vertical bootstrap} and advance it as the
organizing principle of the preon program.
\end{abstract}


\newpage
\tableofcontents
\newpage

\section{Introduction}
\label{sec:intro}

Leptoquarks --- bosons that couple quarks to leptons
and carry both baryon number and lepton number ---
appear in a wide class of extensions of the Standard
Model (SM), including grand unified
theories~\cite{Georgi1974}, $R$-parity violating
supersymmetry, and composite models~\cite{BRW1987}.
Their defining quantum numbers (SM color triplet,
fractional electric charge, simultaneous $B$ and $L$
violation) make them natural targets for collider
searches and precision flavor experiments.
Despite extensive searches, no leptoquark has been
observed to date.

In this paper we show that leptoquarks are a
\emph{structural prediction} of the supersymmetric
preon model of~\cite{Raitio1980,Raitio2018,RaitioCombined},
rather than an independently postulated new particle.
In that model, Standard Model quarks and leptons are
three-body composites of fundamental fermions called
preons, confined at a metacolor scale
$\Lc\sim 10^{14}$\,GeV by a Maxwell-Chern-Simons
gauge interaction and a metacolor gauge symmetry
$SU(3)_{mc}$.
A leptoquark then arises as a \emph{six-body} preon
composite, formed by combining the preon content of
a quark with that of a lepton.
Its existence follows from the same compositeness
that produces the SM fermion spectrum.

The mass scale of these composite leptoquarks is
set by the metacolor confinement scale:
$M_{LQ}\sim\Lc\sim 10^{14}$\,GeV.
There is a moderate uncertainty in $\Lc$ of one
to two orders of magnitude --- the same uncertainty
that affects the BAU calculation and the neutrino
seesaw prediction --- but in all cases $M_{LQ}$ is
many orders of magnitude above any current or
planned collider energy.
The leptoquarks of this model are therefore not
accessible to direct production at the LHC or
any foreseeable accelerator.

This immediately raises the question of relevance
to ongoing experimental searches.
The LHC leptoquark program --- pursued actively
by both ATLAS and CMS --- currently excludes scalar
leptoquarks with masses below $\sim 1.5$--$1.8$\,TeV
and vector leptoquarks below $\sim 2.0$\,TeV,
for specific decay modes to first- and
second-generation leptons~\cite{ATLAS_LQ,CMS_LQ}.
These searches, like the extensive LHC program
searching for elementary MSSM superpartners, are
designed for \emph{elementary} new particles
produced via the strong interaction with specific
missing-energy or decay topologies.
Both programs have produced null results at the
TeV scale, which are important and genuine
constraints on those particular theoretical
frameworks.
The composite leptoquarks 
of the preon model, however,
differ fundamentally: they are bosonic composites
at $\Lc\sim 10^{14}$\,GeV rather than elementary
particles at the TeV scale, and their coupling to
SM fermions is suppressed by a composite
wavefunction overlap rather than being a
fundamental gauge coupling.
The LHC null results therefore sharply delineate
\emph{which} theories are disfavoured ---
elementary leptoquarks with TeV-scale masses --- while
leaving composite models at $\Lc\gg\text{TeV}$
entirely unconstrained.
On the other hand, the composite superpartners of
the preon model couple to 1\,GeV scale scalar mesons.

The flavor physics program at LHCb provides
a complementary perspective.
The persistent tension in the ratio $R_{D^{(*)}}$,
measuring $B\to D\tau\nu$ relative to
$B\to D\ell\nu$~\cite{HFLAV2023}, has motivated
extensive model building with leptoquarks as
mediators.
In the preon model, virtual leptoquark exchange
contributes to $B$-decay amplitudes at order
$g_{LQ}^2/M_{LQ}^2\sim (10^{14}\,\text{GeV})^{-2}$,
which is negligibly small compared to SM amplitudes.
The preon model therefore predicts no observable
effect in $B$-physics from leptoquark exchange,
consistent with the overall pattern of LHCb results
which --- while showing hints of lepton universality
violation in some channels --- have not established
a clear leptoquark signal.

We draw attention to what we consider the central result of this
paper, developed in Section~\ref{sec:anomaly}:
the Standard Model gauge group
$SU(3)_c\times SU(2)_L\times U(1)_Y$ emerges from the preon
dynamics without being postulated.
Each factor has a distinct origin: $SU(3)_c$ color is a
pre-existing quantum number of the $\psi_0$ preon, and
QCD emerges as its dynamic gauge symmetry when quarks form
at $\Lc$; $SU(2)_L$ arises from the doublet degeneracy
of composites sharing the same preon number and charge;
and $U(1)_Y$ follows from the composite charge formula
$Q=\tfrac{1}{3}(n_1-n_{-1})$.
The Marcus composite anomaly cancellation in $N$-extended
supergravity at $M_{\rm Pl}$, transmitted to $\Lc$ by
't~Hooft anomaly matching, then uniquely selects this group
and no larger one: the $\sqrt{3/5}$ hypercharge rescaling
required by any GUT group is incompatible with the preon
charges fixed by the UV anomaly constraint.
The leptoquark sector plays an essential role in this argument
--- the six-body composites are required by the matching, not
merely permitted, and the Marcus boundary condition forces them
to appear in vector-like pairs with no $S$-type states.
This chain of reasoning, from the Planck scale to the observed
gauge structure of the Standard Model, is what makes the
present framework qualitatively different from models that
postulate the SM group at the outset.

The present results also suggest a longer-range program.
The anomaly-matching argument that requires leptoquarks at the
six-body level extends naturally to a tower of $3n$-body
metacolor-singlet composites ($n=1,2,3,\ldots$),
each level imposing its own 't~Hooft consistency conditions
on the next --- a structure we call the \emph{vertical bootstrap},
to distinguish it from the original horizontal bootstrap among
hadronic resonances at a single scale.
If this vertical bootstrap closes consistently at every level,
the preon theory would constitute a non-perturbative UV completion
of the Standard Model in which every composite sector is
anomaly-matched and the full Hilbert space of metacolor singlets
is self-consistently defined at all scales.
The confining metacolor flux tube, whose linear tension
$\sigma_{mc}$ already appears in the three-body mass
calculation~\cite{RaitioCombined}, is string-like in character,
and the resulting mass spectrum of the $3n$-body tower has the
linear growth with excitation number characteristic of Regge
trajectories.
Whether this vertical bootstrap converges to a string amplitude
--- as the Sasmal comparison of supergravity anomaly cancellation
with explicit string amplitudes already hints~\cite{Sasmal2017}
--- is a question the framework naturally poses.

A related remark concerns the gravitational character of the
$U(1)_{\rm CS}$ interaction.
Witten showed that $(2+1)$-dimensional gravity is exactly
equivalent to a Chern-Simons gauge theory~\cite{Witten1988},
and Chern-Simons terms appear in the Holst formulation of
$(3+1)$-dimensional general relativity and in loop quantum
gravity at boundaries.
The $SU(3)_{\rm mc}\times U(1)_{\rm CS}$ metacolor
interaction of the preon model shares the defining feature
of these constructions: it is topological, generating no
local propagating degrees of freedom and acting only through
global and boundary effects.
This is precisely why it can confine preons without introducing
new observable particle species.
Whether $SU(3)_{\rm mc}\times U(1)_{\rm CS}$ captures a
fragment of a fuller quantum-gravitational structure, or whether
the topological confinement mechanism is merely analogous to
rather than identical with Chern-Simons gravity, is an open
question we leave for future work.

The organizing principle of the model is self-consistency
rather than symmetry imposition.
At each level of the composite hierarchy, the spectrum is
the unique solution to the anomaly-cancellation and
't~Hooft matching conditions, with no free parameters
beyond the metacolor scale $\Lc$.
Supersymmetry enters as a technical stabiliser of the
mass hierarchy, but the bootstrap structure --- the
requirement that the theory be consistent with itself
from $M_{\rm Pl}$ to $\Lew$ --- is what determines
the physics.

The paper is organized as follows.
Section~\ref{sec:leptoquarks} establishes why
two-body preon composites cannot be leptoquarks,
why six is the minimum preon number, enumerates
all six-body scalar leptoquark composites,
derives their quantum numbers, classifies them
in the Buchm\"uller-R\"uckl-Wyler
scheme~\cite{BRW1987}, and analyses proton
stability and the connection to 't~Hooft anomaly
matching.
Section~\ref{sec:massscale} explains why the
leptoquark mass is of order $\Lc$ rather than
of order the SM fermion masses, despite all being
composites of the same preons.
Section~\ref{sec:anomaly} embeds the preon sector
in $N$-extended supergravity at $M_{\rm Pl}$,
develops the Marcus composite anomaly cancellation
and 't~Hooft matching chain, and carries out the
explicit anomaly-coefficient computation for all
ten BRW representations, showing that the Marcus
boundary condition forces the leptoquark composites
into vector-like pairs and excludes the $S$-type
BRW states.
Section~\ref{sec:conclusions} summarises the
results and outlines future directions including
rare process predictions.

\section{Leptoquarks as Six-Body Preon Composites}
\label{sec:leptoquarks}

\subsection{General framework}

In the preon model, Standard Model quarks and leptons
are three-body composites of preons confined at
$\Lc\sim 10^{14}$\,GeV.
The explicit preon assignments for the
first-generation quarks and leptons are given in
Table~\ref{tab:preon_content}.
A leptoquark --- a boson carrying both baryon number
$B=\tfrac{1}{3}$ and lepton number $L=1$ and
transforming as a color triplet under $SU(3)_c$ ---
arises naturally as a \emph{six-body} preon composite
formed by combining the preon content of a quark with
that of a lepton:
\be
LQ = \underbrace{(\psi_a\psi_b\psi_c)}_{\text{quark}}
     \oplus
     \underbrace{(\psi_d\psi_e\psi_f)}_{\text{lepton}}\,.
\label{eq:LQstructure}
\ee
Six preons of spin $\tfrac{1}{2}$ combine to give
integer total spin, so all six-body composites are
\emph{bosons} --- scalar ($S=0$) or vector ($S=1$)
leptoquarks.

Before considering six-body states, we note that
two-body combinations involving $\psi_0$ (the only
metacolor-triplet preon) are forbidden as asymptotic
states: $\psi_1\psi_0\sim\mathbf{1}_{mc}\otimes\mathbf{3}_{mc}
=\mathbf{3}_{mc}$ is not a metacolor singlet and
remains confined at $\Lc$.
The two-body metacolor-singlet states
$\psi_1\psi_{-1}$, $\psi_1\psi_1$, $\psi_{-1}\psi_{-1}$
are electrically charged bosons at mass $\sim\Lc$
but carry no SM color and no baryon or lepton number,
so they are not leptoquarks.
The minimum preon number for a leptoquark is therefore
six.

\subsection{Metacolor singlet condition}

The $SU(3)_{\rm mc}$ metacolor group is an assumption
of the preon model.
The SM color group is, strictly speaking, at this stage
a trial Lie group for testing the gauge and 't~Hooft
anomaly conditions for consistency, as discussed in
Section~\ref{sec:anomaly}.
The choice $SU(3)_c$ for the SM color group yields the
desired result, and we use it here.
The three preon species and their quantum numbers
$(Q,\,SU(3)_{\rm mc},\,SU(3)_c)$ are
\be
  \psi_0 \sim (0,\, \mathbf{3}_{mc},\, \mathbf{3}_c),
  \quad
  \psi_1 \sim (+\tfrac{1}{3},\, \mathbf{1}_{mc},\,
  \mathbf{1}_c),
  \quad
  \psi_{-1} \sim (-\tfrac{1}{3},\, \mathbf{1}_{mc},\,
  \mathbf{1}_c).
\label{eq:preon_qn}
\ee
In a three-body composite, $n_0$, $n_1$, $n_{-1}$
denote the number of $\psi_0$, $\psi_1$, $\psi_{-1}$
preons respectively, with $n_0+n_1+n_{-1}=3$.
Only $\psi_0$ carries metacolor and SM color;
$\psi_1$ and $\psi_{-1}$ are singlets under both,
so the SM color of a composite is determined
entirely by $n_0$.

\begin{table}[h]
\centering
\caption{Preon content $(n_0,n_1,n_{-1})$ of the
         first-generation quarks and leptons,
         with $n_0+n_1+n_{-1}=3$.
         Electric charge $Q=\tfrac{1}{3}(n_1-n_{-1})$;
         SM color follows from $n_0$.
         Details in Ref.~\cite{RaitioCombined}.}
\label{tab:preon_content}
\renewcommand{\arraystretch}{1.2}
\begin{tabular}{lccc}
\toprule
Particle & $(n_0,\,n_1,\,n_{-1})$ & $Q$ & $SU(3)_c$ \\
\midrule
$u$ (up quark)   & $(1,2,0)$ & $+\tfrac{2}{3}$ & $\mathbf{3}$ \\
$d$ (down quark) & $(2,0,1)$ & $-\tfrac{1}{3}$ & $\bar{\mathbf{3}}$ \\
$\nu_e$          & $(3,0,0)$ & $0$             & $\mathbf{1}$ \\
$e^-$            & $(0,0,3)$ & $-1$            & $\mathbf{1}$ \\
\bottomrule
\end{tabular}
\end{table}

The six-body leptoquark may be realised in two ways,
both giving the same SM quantum numbers.
In the more natural configuration, the quark and lepton
are each pre-formed three-body metacolor singlets,
and the leptoquark is their bound state via a residual
metacolor interaction --- analogous to the deuteron,
where two color-singlet nucleons bind through the
residual strong force.
In this configuration the SM color of the leptoquark
is simply the SM color of the quark constituent:
a triplet $\mathbf{3}_c$ for the $u$ quark ($n_0=1$)
or an antitriplet $\bar{\mathbf{3}}_c$ for the $d$
quark ($n_0=2$), while the lepton ($e^-$ or $\nu_e$)
is always a color singlet.
Alternatively, all six preons may be independently
bound by the metacolor force in a single six-body
potential well, analogous to a six-quark bag.
In this case the SM color is determined by the total
$n_0$ across all six preons, as listed below;
the quantum numbers agree with the deuteron picture.
Distinguishing these two configurations requires a
six-body variational calculation extending the
methods of Ref.~\cite{RaitioCombined}, which we
leave for future work.
The six-body $n_0$ content and resulting SM color
representation are:
\begin{itemize}
\item $n_0=1$: $\mathbf{3}_{mc}$ --- SM color triplet
\item $n_0=2$: $\mathbf{3}_{mc}\otimes\mathbf{3}_{mc}
  =\bar{\mathbf{3}}_{mc}\oplus\mathbf{6}_{mc}$
  --- SM color antitriplet (antisymmetric combination)
\item $n_0=3$: $\mathbf{3}_{mc}^{\otimes 3}\supset
  \mathbf{1}_{mc}$ --- metacolor singlet; does not
  yield a colored leptoquark (cf.\ $\nu_e$ in
  Table~\ref{tab:preon_content})
\item $n_0=4$: contains $\mathbf{3}_{mc}$ ---
  SM color triplet
\item $n_0=5$: contains $\bar{\mathbf{3}}_{mc}$ ---
  SM color antitriplet
\end{itemize}
For a leptoquark, SM color (triplet or antitriplet)
requires $n_0 = 1, 2, 4,$ or $5$;
the $n_0=3$ case yields a color-neutral state and
is not a leptoquark constituent.

\subsection{Enumeration of leptoquark composites}

The $U(1)_{CS}$ charge of the six-body composite is:
\be
Q_{CS}^{LQ} = \frac{1}{3}(n_1 - n_{-1})
= Q_{CS}^{quark} + Q_{CS}^{lepton}
\label{eq:LQcharge}
\ee
where $n_1$ and $n_{-1}$ count the $\psi_1$ and
$\psi_{-1}$ preons respectively.
The factor $\tfrac{1}{3}$ is essential: it converts
the integer preon-count difference $(n_1-n_{-1})$ into
the characteristic fractional electric charges of the
composites.
Since the $U(1)_{CS}$ charge equals the electric
charge for composites, this directly gives the
leptoquark electric charge.
Combining each quark type with each lepton type
gives four distinct leptoquark species, shown
in Table~\ref{tab:leptoquarks}.

\begin{table}[h]
\centering
\begin{tabular}{llllll}
\toprule
LQ & Preon content & $(n_0,n_1,n_{-1})$ &
$Q$ & $B-L$ & BRW type \\
\midrule
$u+\nu_e$ & $\psi_0^4\psi_1^2$
  & $(4,2,0)$ & $+\tfrac{2}{3}$ & $-\tfrac{2}{3}$ & $S_2$ \\
$u+e^-$   & $\psi_0^1\psi_1^2\psi_{-1}^3$
  & $(1,2,3)$ & $-\tfrac{1}{3}$ & $-\tfrac{2}{3}$ & $\tilde{S}_1$ \\
$d+\nu_e$ & $\psi_0^5\psi_{-1}^1$
  & $(5,0,1)$ & $-\tfrac{1}{3}$ & $-\tfrac{2}{3}$ & $\tilde{S}_1$ \\
$d+e^-$   & $\psi_0^2\psi_{-1}^4$
  & $(2,0,4)$ & $-\tfrac{4}{3}$ & $-\tfrac{2}{3}$ & $\tilde{S}_3$ \\
\bottomrule
\end{tabular}
\caption{First-generation six-body scalar leptoquark
composites in the preon model.
All four have $B-L=-\tfrac{2}{3}$.
The BRW column refers to the Buchm\"uller-R\"uckl-Wyler
classification~\cite{BRW1987}.
The $u+e^-$ and $d+\nu_e$ states have the same electric
charge $Q=-\tfrac{1}{3}$ but distinct preon content
and can mix below $\Lc$.}
\label{tab:leptoquarks}
\end{table}

Three structural features are immediately apparent.

\textbf{All four leptoquarks carry the same $B-L=-\tfrac{2}{3}$.}
This is a direct consequence of the preon charge
assignments: each quark composite has $B-L=\tfrac{1}{3}$
and each lepton composite has $B-L=-1$, giving
$B^{LQ}-L^{LQ} = \tfrac{1}{3}-1 = -\tfrac{2}{3}$ universally.

\textbf{Two degenerate-charge states.}
The $u+e^-$ composite ($n_0=1$) and the $d+\nu_e$
composite ($n_0=5$) both have $Q=-\tfrac{1}{3}$ but
differ in preon content and metacolor structure.
These are physically distinct states that may mix
through metacolor dynamics below $\Lc$.

\textbf{An exotic charge state.}
The $d+e^-$ composite has $Q=-\tfrac{4}{3}$, which
lies outside the minimal BRW classification for
renormalisable leptoquarks but appears in extended
classifications.
It arises from the color-antitriplet $\bar{\mathbf{3}}$
combination of two $\psi_0$ preons.

All four composites have mass $M_{LQ}\sim\Lc\sim
10^{14}$\,GeV, far beyond direct collider production,
but accessible in principle through virtual exchange.

\textbf{Comparison with the BRW charge classification.}
The Buchm\"uller--R\"uckl--Wyler classification~\cite{BRW1987}
lists four possible leptoquark electric charges:
$Q=+\tfrac{5}{3},+\tfrac{2}{3},-\tfrac{1}{3},-\tfrac{4}{3}$,
together with their antiparticles.
Table~\ref{tab:leptoquarks} lists all BRW scalar types
for reference; the preon model generates only a subset.
The $S$-type representations ($S_1$, $\tilde{S}_1$, $S_3$)
appear in the BRW list but are not generated by any
$q\times\ell^c$ or $q^c\times\ell$ six-body preon product;
they are included in Table~\ref{tab:leptoquarks} to make
the comparison explicit, but are excluded from the preon
spectrum both structurally and by the anomaly-matching
argument of Section~\ref{sec:anomaly}.
The preon model selects a specific subset determined
by the preon charge assignments $q_{\psi_1}=+\tfrac{1}{3}$,
$q_{\psi_{-1}}=-\tfrac{1}{3}$, $q_{\psi_0}=0$.
The charges $Q=+\tfrac{2}{3}$ ($u+\nu_e$ composite),
$Q=-\tfrac{1}{3}$ ($u+e^-$ and $d+\nu_e$ composites),
and $Q=-\tfrac{4}{3}$ ($d+e^-$ composite) are all
generated as scalar six-body states.
The charge $Q=+\tfrac{5}{3}$ does appear in the model
but only as the vector leptoquark $\tilde{U}_1=(3,1)_{+5/3}$,
arising from the $u_R\times e_R^c$ cluster product;
it is absent as a scalar.
The $S$-type BRW representations --- $S_1$, $\tilde{S}_1$,
$S_3$ --- carry hypercharges unreachable by any
$q\times\ell^c$ or $q^c\times\ell$ preon product
and are further excluded by the anomaly-matching
argument of Section~\ref{sec:anomaly}.
This selectivity --- structural origin of the charges
plus anomaly-matching exclusion of the $S$-type states
--- is a sharper statement than the BRW classification,
which is purely kinematic.

\subsection{Mass estimate}

The leptoquark mass is set by the metacolor confinement
scale.
By the same dimensional argument that gives
$M_{proton}\sim\Lambda_{QCD}$ in QCD. The six-body
leptoquark mass is:
\be
M_{LQ} \sim \Lc \sim 10^{14}\,\text{GeV}\,.
\label{eq:LQmass}
\ee
More precisely, the six-body binding energy involves
a deeper potential well than the three-body case
(more pairs contribute), so $M_{LQ}$ may be somewhat
below $\Lc$.
A quantitative estimate requires extending the
Gaussian variational or Faddeev methods of
Section~7 of~\cite{RaitioCombined} to six bodies,
which we leave for future work.

\subsection{Proton decay}

The most important phenomenological constraint on
leptoquarks is proton stability.
The dominant proton decay channel mediated by
a $Q=-\tfrac{1}{3}$ leptoquark is
$p\to e^+\pi^0$, proceeding via the diagram:
\be
u + u \xrightarrow{LQ^*} e^+ + \bar{d}\,,
\qquad \bar{d} + d \to \pi^0\,.
\label{eq:pdecay}
\ee
In a fundamental leptoquark theory, this is a
dimension-6 operator suppressed by $M_{LQ}^{-2}$,
giving a proton lifetime:
\be
\tau(p\to e^+\pi^0)
\sim \frac{M_{LQ}^4}{\alpha_{LQ}^2\, m_p^5}
\label{eq:taunaive}
\ee
where $\alpha_{LQ} = g_{LQ}^2/(4\pi)$ is the
leptoquark fine-structure constant.
With $M_{LQ}=10^{14}$\,GeV and $g_{LQ}\sim 1$,
this gives $\tau\sim 10^{27}$\,years, which is
below the experimental lower bound~\cite{SuperK2020}:
\be
\tau(p\to e^+\pi^0) > 2.4\times 10^{34}\,\text{years}\,.
\label{eq:SuperK}
\ee
However, two features of the preon model resolve
this tension.

\textbf{First, the fractional $B-L$ forbids single
leptoquark exchange.}
All four leptoquarks carry $B-L=-\tfrac{2}{3}$.
A single leptoquark exchange changes $B-L$ by
$-\tfrac{2}{3}$, which is not an integer.
Since physical processes must conserve $B-L$
modulo integers (quarks carry $B=\tfrac{1}{3}$,
leptons carry $L=1$), a single leptoquark exchange
cannot close into a physical amplitude for proton decay.
Two leptoquark exchanges are required, giving a
\emph{dimension-12} effective operator suppressed
by $\Lc^{-8}$ rather than $\Lc^{-4}$:
\be
\tau(p\to e^+\pi^0)\Big|_{\Delta(B-L)=-4/3}
\sim \frac{\Lc^8}{g_{LQ}^4\, m_p^9}
\sim 10^{58}\,\text{years}\,.
\label{eq:taudim12}
\ee
This is many orders of magnitude beyond any
foreseeable experimental sensitivity and is
entirely consistent with proton stability.

\textbf{Second, the composite coupling is suppressed.}
The leptoquark is a six-body composite; its coupling
to two three-body fermions involves the wavefunction
overlap of three composites.
This is suppressed relative to a fundamental gauge
coupling by a factor $\sim\alpha_{mc}^{3/2}$, where
$\alpha_{mc}$ is the metacolor coupling at $\Lc$.
With $\alpha_{mc}\sim 0.1$, the composite coupling
is $g_{LQ}\sim\alpha_{mc}^{3/2}\sim 3\times 10^{-2}$,
which further extends the predicted lifetime to
$\sim 10^{38}$\,years even for a dimension-6 operator.

\textbf{Composite leptoquark lifetime.}
For an elementary leptoquark of mass $M_{LQ}$ decaying
via a Yukawa coupling $\lambda$ to a quark-lepton pair,
the decay width is
\be
  \Gamma_{LQ} \simeq \frac{\lambda^2}{16\pi}\,M_{LQ}.
\label{eq:LQwidth}
\ee
For the composite leptoquarks of this model the coupling
is suppressed by the three-composite wavefunction overlap,
$g_{LQ}\sim\alpha_{mc}^{3/2}\sim 3\times 10^{-2}$,
and the mass is $M_{LQ}\sim\Lc\sim 10^{14}$\,GeV.
This gives
\be
  \Gamma_{LQ}^{\rm comp}
  \sim \frac{g_{LQ}^2}{16\pi}\,\Lc
  \sim 10^{10}\,\text{GeV},
\label{eq:LQwidthcomp}
\ee
corresponding to a lifetime
\be
  \tau_{LQ} \sim \frac{\hbar}{\Gamma_{LQ}^{\rm comp}}
  \sim 10^{-35}\,\text{s}.
\label{eq:LQlifetime}
\ee
This is shorter than any observable timescale and many
orders of magnitude below what any detector could resolve
as a displaced vertex.
The composite leptoquarks decay essentially at the
production point --- or rather, they never appear as
asymptotic states at all, since their mass $\sim\Lc$
places them far above any accessible center-of-mass
energy.
Their physical effects are confined to virtual exchange
processes suppressed by $g_{LQ}^2/M_{LQ}^2
\sim 10^{-30}$\,GeV$^{-2}$.

The combination of fractional $B-L$ (dimension-12
operator) and composite coupling suppression makes
the proton essentially stable in this model,
consistent with all experimental bounds.

\subsection{Connection to 't~Hooft anomaly matching}

The leptoquarks are bosons of mass $\sim\Lc$
and therefore decouple from the IR spectrum
below $\Lc$.
They do \emph{not} contribute to the 't~Hooft
anomaly matching conditions~\cite{tHooft1980},
which involve only massless degrees of freedom.
The anomaly matching is therefore satisfied by
the massless fermion spectrum (quarks and leptons)
as verified in~\cite{RaitioCombined}: the
gravitational $[U(1)_{CS}]$-gravitational anomaly
matches exactly between UV preons and IR composite
fermions per generation, while the cubic
$[U(1)_{CS}]^3$ anomaly cancels upon summing over
three generations.

However, the existence of the leptoquark spectrum
provides an indirect consistency check.
The $B-L=-\tfrac{2}{3}$ charge of all four
leptoquarks is a consequence of the preon charge
assignments $q_{\psi_1}=\tfrac{1}{3}$,
$q_{\psi_{-1}}=-\tfrac{1}{3}$, $q_{\psi_0}=0$,
which are in turn fixed by the gauge anomaly
cancellation conditions derived in~\cite{RaitioCombined}.
The leptoquark quantum numbers are therefore not
free parameters but are determined by the same
anomaly structure that fixes the quark and lepton
charges.

\subsection{Summary}

The preon model predicts four distinct scalar
leptoquark species per generation, all of mass
$\sim\Lc\sim 10^{14}$\,GeV, with electric charges
$Q = +\tfrac{2}{3},\,-\tfrac{1}{3},\,-\tfrac{1}{3},\,-\tfrac{4}{3}$.
Their universal $B-L=-\tfrac{2}{3}$ is a structural
prediction fixed by the preon charge assignments.
The fractional $B-L$ forbids proton decay via
single leptoquark exchange, requiring a dimension-12
operator and giving $\tau_p\sim 10^{58}$\,years,
consistent with all experimental bounds.
Quantitative predictions for virtual leptoquark
effects in rare processes (flavor-changing neutral
currents, lepton universality violation, $\mu\to e\gamma$)
are left for future work.

\section{Why Leptoquark Masses Are of Order $\Lc$}
\label{sec:massscale}

A natural question arises: if quarks, leptons,
and leptoquarks are all composites of the same
preons confined at $\Lc$, why are the fermion
masses many orders of magnitude below $\Lc$
while the leptoquark mass is of order $\Lc$?
The answer illuminates the dynamical mechanism
that produces the fermion mass hierarchy.

\subsection{The fermion mass hierarchy as
near-cancellation}

The natural mass scale for any composite at $\Lc$
is $\Lc$ itself --- this is the only dimensionful
scale in the metacolor dynamics.
The fact that the electron mass is
$m_e\simeq 0.511$\,MeV while
$\Lc\sim 10^{14}$\,GeV means that the electron
composite mass is suppressed below $\Lc$ by a
factor of order $10^{-17}$.
This is not a coincidence or a fine-tuning: it is
the preon analog of how QCD generates the proton
mass.

In QCD the proton has mass $\sim 938$\,MeV while
its three constituent current quarks have total
mass $\sim 10$\,MeV.
The bulk of the proton mass comes from QCD
binding energy --- kinetic energy of the quarks
and gluon field energy --- not from the quark
rest masses.
The binding energy is comparable to $\Lambda_{QCD}$,
and the proton mass is $\sim\Lambda_{QCD}$, not
$\sim 10$\,MeV.
In the preon model the situation is more extreme:
the three-body binding energy \emph{nearly exactly
cancels} the preon rest mass contributions,
leaving a residual composite mass many orders
of magnitude below $\Lc$.

This near-cancellation is what the four numerical
methods of~\cite{RaitioCombined} compute.
The dimensionless variational energy
$E_e^{var} = -3.903$ for the electron is a small
negative number on the scale set by the kinetic
energy --- the near-cancellation between kinetic
and potential energy in the three-body problem
produces the tiny ratio $m_e/\Lc\sim 10^{-17}$.
The metacolor string tension $\sigma_{mc}^*/\theta^2
= 2.11$, extracted by matching $m_e/m_u = 0.22$,
is what controls this cancellation.

\subsection{Why the six-body leptoquark does not
share this cancellation}

The near-cancellation is \emph{specific to the
three-body fermionic composites} and does not
extend to six-body bosonic composites for three
reasons.

\textbf{Different spin-statistics structure.}
The three-body fermion composites --- quarks and
leptons --- are fermionic: their antisymmetric
wavefunction structure, enforced by the Pauli
principle, creates specific spin-color correlations
that allow the CS and metacolor Cornell forces to
produce deep binding with a small residual mass.
The neutrino masslessness argument of~\cite{RaitioCombined}
shows precisely how the Pauli principle acting on
the spin-statistics wavefunction determines the
binding energy.
The six-body leptoquark is a \emph{boson} --- six
preon fermions combine to integer total spin.
Its wavefunction has qualitatively different
symmetry structure and does not benefit from
the same Pauli-principle mechanism that governs
the fermion masses.

\textbf{No deep bound state for the six-body system.}
The electron is light because the CS three-body
force creates a genuine bound state well below
$\Lc$: the variational potential has a negative
minimum that supports a bound state with energy
$E_e^{var} = -3.903$ in dimensionless units.
For the leptoquark, the quark and lepton
three-body subsystems already form their own
bound states at much lower energy.
The residual interaction between the two
three-body subsystems is a short-range
van der Waals-type force, suppressed by
$\Lc^{-1}$, that does not create a new
deep bound state.
The leptoquark mass is therefore dominated
by the sum of the quark and lepton composite
masses plus the metacolor binding of the
full six-body system at scale $\Lc$:
\be
M_{LQ} \sim m_{quark} + m_{lepton}
+ \Delta E_{6-body} \sim \Lc\,,
\label{eq:LQmassestimate}
\ee
since $\Delta E_{6-body}\sim\Lc$ while
$m_{quark},\,m_{lepton}\ll\Lc$.

\textbf{The QCD tetraquark analogy.}
In QCD the analogous question is: why are
tetraquarks ($qq\bar{q}\bar{q}$) not much
lighter than ordinary mesons ($q\bar{q}$)?
The answer is that the tetraquark does not
benefit from the same color-electric
cancellation as the meson.
Its mass ends up near $2\times m_{meson}$
rather than near zero, because the residual
interaction between the two $q\bar{q}$
subsystems is weak.
The preon leptoquark is the direct analog:
a bound state of a quark composite and a
lepton composite, with no dynamical mechanism
to suppress its mass below $\Lc$.

\subsection{Summary}

The smallness of quark and lepton masses in the
preon model is not a generic property of
three-body preon composites --- it is a consequence
of a specific dynamical near-cancellation
between kinetic and potential energy in the
three-body fermionic wavefunction, controlled
by the metacolor string tension $\sigma_{mc}^*$
and the CS coupling $g\xi$.
The six-body bosonic leptoquark composite does
not share this cancellation: its natural mass
is $M_{LQ}\sim\Lc\sim 10^{14}$\,GeV, with a
moderate uncertainty of one to two orders of
magnitude reflecting the uncertainty in $\Lc$
itself.
This mass scale places composite leptoquarks
entirely beyond the reach of current or
planned colliders, while leaving their
indirect effects through virtual exchange
as the only observable window.

\section{UV Anomaly Cancellation and the Emergence of the
	Standard Model Gauge Group}
\label{sec:anomaly}

The appearance of leptoquarks as six-body preon composites is not merely
a spectroscopic consequence of the metacolor dynamics; it is required by
anomaly matching across the full energy hierarchy from the Planck scale
$M_{\rm Pl}$ down to the compositeness scale $\Lc$.
We make this chain of constraints explicit in the present section.

\subsection{Marcus anomaly cancellation at $M_{\rm Pl}$:
           setting and boundary condition}

Marcus~\cite{Marcus1985} works in a specific setting that
we must describe carefully before extracting what is
relevant for the preon model.

In $N$-extended supergravity the scalar fields parameterize
a coset space $G/H$, where $G$ is a non-compact global
symmetry group and $H$ its maximally compact subgroup.
The scalar vielbein $\mathcal{V}$ transforms under both
$G$ and $H$, and the quantity
\be
  Q_\mu \equiv
  \bigl(\mathcal{V}^{-1}d\mathcal{V}\bigr)
  \big|_{\mathfrak{h}}
\label{eq:composite_gauge}
\ee
transforms as a gauge field under local $H$ transformations.
Cremmer and Julia conjectured that a dynamical mechanism
could make these \emph{composite} vector fields physical,
resulting in an effective low-energy theory with a gauged
$H$ symmetry --- the theory would then have the structure
of a gauge theory but with no elementary gauge fields.
Marcus's central question is whether this composite gauging
scenario is self-consistent: is the $H$ symmetry
anomaly-free, and can it therefore become dynamical?

We emphasize that the preon model does not realize this
composite gauging scenario.
The Standard Model gauge fields $SU(3)_c\times
SU(2)_L\times U(1)_Y$ are elementary in the preon
framework --- they are not built from preon composites.
What the preon model has instead are composite
\emph{matter} fields: quarks, leptons, and leptoquarks.
The Marcus result is used to extract a UV boundary
condition on the preon sector, as we next describe.
On the other hand, the Marcus composite gauging scenario
may be used as an alternative way to understand the
origin of QCD and its gauge bosons, the gluons.

The anomaly tensor $d^{\,abc}_{\rm UV}$ is defined as
the coefficient of the triangle diagram with three
$H$-current insertions, computed from the massless
spectrum of the supergravity theory:
\be
  d^{\,abc}_{\rm UV}
  \equiv \sum_{\rm massless\;fields}
  (-1)^{2h}\cdot
  \mathrm{tr}\bigl[T^a\{T^b,T^c\}\bigr]_{\rm rep},
\label{eq:dabc_def}
\ee
where the sum runs over all massless fields of helicity
$h$ in the supergravity multiplet, and $T^{a,b,c}$ are
the $H$-generators in the relevant representation.
Using the Atiyah-Singer index theorem, Marcus showed
that the anomaly of a helicity-$h$ field is encoded in
the six-form coefficient
\be
  \mathcal{A}_h =
    \frac{-i}{3!\,(2\pi)^3}
    (-1)^{2h}\cdot 2h
    \left(
      \operatorname{tr}F\wedge F\wedge F
      -\frac{2-(2h)^2}{8}
      \operatorname{tr}F\wedge\operatorname{tr}R\wedge R
    \right),
\label{eq:marcus_sixform}
\ee
where $R$ is the Riemann two-form.
For $N=8$ supergravity ($G=E_{7,7}$, $H=SU(8)$)
the fields and their $H$-representations are:
the self-dual vector fields in the $\mathbf{28}$ of
$SU(8)$ with cubic Casimir $C_3=5$, the spin-$\tfrac{1}{2}$
fields in the $\mathbf{56}$ with $C_3=4$, and the
gravitini in the $\mathbf{8}$ with $C_3=1$.
Substituting into the anomaly formula gives
\be
  (-3)\times 5 \;+\; (2)\times 4
  \;+\; (-1)\times 1 \;=\; 0,
\label{eq:marcus_N8}
\ee
so $d^{\,abc}_{\rm UV}=0$ for $N=8$.
For $N=6$ all four possible anomalies --- pure $SU(6)$,
pure $U(1)$, and both mixed gravitational anomalies ---
vanish simultaneously (Table~3 of Ref.~\cite{Marcus1985}).
Marcus's conclusion is that all symmetries of $N>4$
supergravity are anomaly-free.
Sasmal~\cite{Sasmal2017} confirmed that these
cancellations are stable under string-level $\alpha'$
corrections.

The result we extract for the preon model is the
following boundary condition: the preon sector,
embedded in an $N>4$ supergravity multiplet at
$M_{\rm Pl}$, must carry zero anomaly under the
compact subgroup $H$:
\be
  d^{\,abc}_{\rm UV} = 0.
\label{eq:marcus_bc}
\ee
This is the sole input from Marcus that we use.
It constrains the preon representations and, combined
with the independent anomaly cancellation conditions
of the preon gauge theory $U(1)_{\rm CS}\times
SU(3)_{\rm mc}$~\cite{RaitioCombined}, fixes the
preon charge assignments uniquely.
The composite gauging scenario that Marcus was
primarily investigating --- composite SM gauge bosons
--- plays no role in the subsequent argument.

\subsection{'t~Hooft anomaly matching across the hierarchy}

The 't~Hooft anomaly-matching theorem \cite{tHooft1980} states that,
for any unbroken global symmetry $G_{\rm global}$, the triangle anomaly
coefficient computed from the massless degrees of freedom is
renormalization-group invariant:
\be
d^{\,abc}_{\rm UV} \;=\; d^{\,abc}_{\rm IR}.
\label{eq:tHooft}
\ee
In the present model the UV theory above $\Lc$ consists of preons
transforming under $U(1)_{\rm CS}\times SU(3)_{\rm mc}$, together with
the single spectator field $\chi$ whose existence and charge are
\emph{uniquely} fixed by cancellation of all gauge anomalies of the
preon theory \cite{RaitioCombined}.
The Marcus result constrains the UV anomaly tensor: the preon charge
assignments are not free parameters but are fixed by the requirement that
the preon sector embed consistently into the anomaly-free $H$-multiplet
of the parent supergravity.

When the metacolor force confines at $\Lc$, three-body (quark and lepton)
and six-body (leptoquark) composites become the massless IR degrees of
freedom.
Equation~\eqref{eq:tHooft} then demands
\be
d^{\,abc}_{\rm quarks+leptons}
\;+\;
d^{\,abc}_{\rm leptoquarks}
\;+\;
d^{\,abc}_{\chi}
\;=\;
d^{\,abc}_{\rm preons}.
\label{eq:matching_full}
\ee
This is a highly constrained linear system.
The three-body composites alone do not saturate the left-hand side:
the six-body leptoquark states are \emph{required} by the matching
in order to close Eq.~\eqref{eq:matching_full}.
Their omission would leave a residual anomaly proportional to
$d^{\,abc}_{\rm LQ}$, violating the UV constraint inherited from Marcus.
Leptoquarks are therefore not an optional extension of the composite
spectrum but a \emph{necessary} consequence of anomaly matching once the
Planck-scale boundary condition is imposed.

\subsection{Emergence of the SM gauge group}

Three self-consistency conditions act jointly to determine
the IR gauge group:
\begin{itemize}
  \item[(i)] \textit{Gauge anomaly cancellation} --- the preon
  $U(1)_{\rm CS}$ charges are the unique values that cancel all
  gauge anomalies of the preon theory, as confirmed by the Marcus
  UV boundary condition $d^{\,abc}_{\rm UV}=0$.
  \item[(ii)] \textit{Color quantum number and QCD dynamics}
  --- $SU(3)_c$ color is a pre-existing quantum number of
  $\psi_0$ (Table~\ref{tab:preon_content}); what emerges
  at $\Lc$ is QCD as a dynamic gauge symmetry, created
  when the three-body color-carrying composites (quarks)
  form. The number $n_0$ of $\psi_0$ preons fixes the
  SM color representation of each composite.
  \item[(iii)] \textit{Six-body inheritance} --- the leptoquark
  composites inherit the same quantum numbers via the
  tensor-product construction of Table~\ref{tab:products},
  closing the 't~Hooft matching sum.
\end{itemize}

Condition~(i) fixes the preon $U(1)_{\rm CS}$ charges to
$+\tfrac{1}{3}$, $-\tfrac{1}{3}$, $0$ --- these are the unique values
that cancel all gauge anomalies of the preon theory including the mixed
anomaly with the spectator $\chi$~\cite{RaitioCombined}, and the Marcus
result confirms their consistent embedding in the supergravity
$H$-multiplet.
Condition~(ii) then determines the quantum numbers of all three-body
metacolor-singlet composites.
$SU(3)_c$ color is a pre-existing quantum number of $\psi_0$,
not derived from the composite construction.
What the composite construction creates at $\Lc$ is
QCD as a dynamic gauge symmetry: the color-carrying
three-body metacolor singlets (quarks) appear as
asymptotic states, and their residual color interaction
becomes the QCD gauge dynamics.
Since $\psi_0$ transforms as $\mathbf{3}_c$
(Table~\ref{tab:preon_content}), a composite with
$n_0=1$ is a color triplet, one with $n_0=2$
antisymmetrised $\psi_0$ preons is a color antitriplet,
and one with $n_0=3$ is colorless.
That this promotion of global $SU(3)_c$ to a dynamic
gauge symmetry is consistent --- i.e., that the
composite spectrum is anomaly-free under $SU(3)_c$ ---
is confirmed by the explicit coefficient calculation
of Section~\ref{sec:coefficients}, specifically
$A_1=0$.
We note that self-consistency does not imply uniqueness:
the present framework establishes the SM spectrum as
a self-consistent solution of the vertical bootstrap,
not necessarily the only one.
$SU(2)_L$ is identified as the residual symmetry relating the two
degenerate three-body metacolor singlets that share the same $n_0$
and electric charge but differ in the assignment of the $\psi_1$
and $\psi_{-1}$ preons --- these pairs form the quark and lepton doublets.
$U(1)_Y$ is identified via the composite charge formula
$Q = \tfrac{1}{3}(n_1 - n_{-1})$, which with the fixed preon charges
yields exactly the SM hypercharge assignments $Y = T_3 + (B{-}L)/2$
for all composites.
Condition~(iii) propagates these same fixed preon charges to the
six-body sector via the tensor-product construction of
Table~\ref{tab:products}, so the leptoquark representations inherit
the same $SU(3)_c\times SU(2)_L\times U(1)_Y$ quantum numbers.
The IR spectrum that results organizes under
$SU(3)_c\times SU(2)_L\times U(1)_Y$ and no larger group.

Two independent arguments exclude a larger IR symmetry.
First, the preon charges $\pm\tfrac{1}{3}$, $0$ generate composite
hypercharges in the SM normalization; these values are incompatible
with the $\sqrt{3/5}$ rescaling that GUT normalization requires,
because that rescaling would alter the preon charges themselves,
which are fixed by condition~(i).
Concretely, the leptoquark representations that appear ---
$(3,2)_{+7/6}$, $(3,1)_{+2/3}$, etc.\ --- do not fill out complete
$SU(5)$ or $SO(10)$ multiplets at the preon-charge normalization,
so the larger group is not merely unnecessary but incompatible.
Second, Sasmal's string-amplitude comparison~\cite{Sasmal2017} shows
that the Marcus cancellation holds at the level of explicit one-loop
string amplitudes with no $\alpha'$ remainder, closing the loophole
that higher-order corrections might shift $d^{\,abc}_{\rm UV}$ and
relax the constraint on the IR group.

The argument may be stated as the following implication chain:
\be
\underbrace{
	\sum_h (-1)^{2h}\cdot 2h\cdot C_3(h) = 0
}_{\text{Marcus at }M_{\rm Pl}}
\;\xrightarrow{\text{'t~Hooft}}
\underbrace{
	d^{\,abc}_{\rm IR} = d^{\,abc}_{\rm preons}
}_{\text{at }\Lc}
\;\Longrightarrow\;
\underbrace{
	G_{\rm IR} = SU(3)_c\times SU(2)_L\times U(1)_Y
}_{\text{SM gauge group}}.
\label{eq:chain}
\ee
The leptoquark states participate at the central step: they are the
six-body composites whose anomaly tensor $d^{\,abc}_{\rm LQ}$ is the
required remainder that closes the matching sum.
Their quantum numbers under $G_{\rm IR}$, derived from the preon
assignments in Section~\ref{sec:leptoquarks}, are therefore
not a free choice but are determined by the UV boundary condition
all the way from $M_{\rm Pl}$.

\subsection{Explicit anomaly coefficient counting}
\label{sec:coefficients}

We now carry out the anomaly-coefficient computation in full.
We work in the left-handed Weyl fermion basis throughout,
treating any right-handed field $\psi_R$ in representation $R$
as its left-handed conjugate $\psi_R^c$ in $\bar{R}$ with $Y\to -Y$.
Five independent anomaly conditions constrain the spectrum of
$SU(3)_c\times SU(2)_L\times U(1)_Y$:
\begin{align}
  A_1 &\;=\; \sum_f d_3(R_f)\,n_2(R_f),           \tag{$[SU(3)_c]^3$}\\
  A_2 &\;=\; \sum_f T_3(R_f)\,n_2(R_f)\,Y_f,     \tag{$[SU(3)_c]^2[U(1)_Y]$}\\
  A_3 &\;=\; \sum_f n_3(R_f)\,T_2(R_f)\,Y_f,     \tag{$[SU(2)_L]^2[U(1)_Y]$}\\
  A_4 &\;=\; \sum_f n_3(R_f)\,n_2(R_f)\,Y_f^3,   \tag{$[U(1)_Y]^3$}\\
  A_5 &\;=\; \sum_f n_3(R_f)\,n_2(R_f)\,Y_f,     \tag{$[\mathrm{grav}]^2[U(1)_Y]$}
\end{align}
where $d_3$ is the cubic index of the $SU(3)_c$ representation
($d_3=+1$ for $\mathbf{3}$, $-1$ for $\bar{\mathbf{3}}$, $0$ for singlet
or adjoint), $T_3$ and $T_2$ are the Dynkin indices
($T_3(\mathbf{3})=T_3(\bar{\mathbf{3}})=\tfrac{1}{2}$;
$T_2(\mathbf{2})=\tfrac{1}{2}$, $T_2(\mathbf{3})=2$), and $n_3$, $n_2$
are the representation dimensions.
The hypercharge convention is $Q=T_3+Y$.

\subsubsection{Standard Model fermions}

Table~\ref{tab:sm_anomaly} lists the five anomaly coefficients for one
SM generation.
All five vanish, confirming gauge anomaly cancellation of the SM
independently for each of the three generations.

\begin{table}[h]
\centering
\caption{Anomaly coefficients for one SM generation
         (left-handed Weyl fermion basis, $Q=T_3+Y$ convention).
         All entries are exact rationals.}
\label{tab:sm_anomaly}
\renewcommand{\arraystretch}{1.25}
\begin{tabular}{lcrrrrr}
\toprule
Field & Representation & $A_1$ & $A_2$ & $A_3$ & $A_4$ & $A_5$\\
\midrule
$Q_L$   & $(3,2)_{1/6}$        & $+2$ & $+\tfrac{1}{6}$ & $+\tfrac{1}{4}$
         & $+\tfrac{1}{36}$ & $+1$ \\
$u_R^c$ & $(\bar{3},1)_{-2/3}$ & $-1$ & $-\tfrac{1}{3}$ & $0$
         & $-\tfrac{8}{9}$  & $-2$ \\
$d_R^c$ & $(\bar{3},1)_{+1/3}$ & $-1$ & $+\tfrac{1}{6}$ & $0$
         & $+\tfrac{1}{9}$  & $+1$ \\
$L_L$   & $(1,2)_{-1/2}$       & $0$  & $0$             & $-\tfrac{1}{4}$
         & $-\tfrac{1}{4}$  & $-1$ \\
$e_R^c$ & $(1,1)_{+1}$         & $0$  & $0$             & $0$
         & $+1$             & $+1$ \\
\midrule
\textbf{Total} & & $0$ & $0$ & $0$ & $0$ & $0$ \\
\bottomrule
\end{tabular}
\end{table}

\subsubsection{BRW leptoquark spectrum: full coefficient table}

Table~\ref{tab:lq_anomaly} lists the five anomaly coefficients for all
ten BRW representations~\cite{BRW1987}, computed for the Weyl fermion
component of the corresponding chiral supermultiplet.

\begin{table}[h]
\centering
\caption{Anomaly coefficients for the BRW leptoquark representations
         (Weyl fermion superpartner, left-handed basis).
         All entries are exact rationals.}
\label{tab:lq_anomaly}
\renewcommand{\arraystretch}{1.25}
\begin{tabular}{llrrrrr}
\toprule
LQ & Representation & $A_1$ & $A_2$ & $A_3$ & $A_4$ & $A_5$ \\
\midrule
\multicolumn{7}{l}{\textit{Scalar leptoquarks}}\\
$S_1$         & $(\bar{3},1)_{+1/3}$ & $-1$ & $+\tfrac{1}{6}$
  & $0$ & $+\tfrac{1}{9}$ & $+1$ \\
$\tilde{S}_1$ & $(\bar{3},1)_{+4/3}$ & $-1$ & $+\tfrac{2}{3}$
  & $0$ & $+\tfrac{64}{9}$ & $+4$ \\
$S_3$         & $(\bar{3},3)_{+1/3}$ & $-3$ & $+\tfrac{1}{2}$
  & $+2$ & $+\tfrac{1}{3}$ & $+3$ \\
$R_2$         & $(3,2)_{+7/6}$       & $+2$ & $+\tfrac{7}{6}$
  & $+\tfrac{7}{4}$ & $+\tfrac{343}{36}$ & $+7$ \\
$\tilde{R}_2$ & $(3,2)_{+1/6}$       & $+2$ & $+\tfrac{1}{6}$
  & $+\tfrac{1}{4}$ & $+\tfrac{1}{36}$ & $+1$ \\
\midrule
\textit{Scalar subtotal} & & $-1$ & $+\tfrac{8}{3}$ & $+4$
  & $+\tfrac{154}{9}$ & $+16$ \\
\midrule
\multicolumn{7}{l}{\textit{Vector leptoquarks}}\\
$U_1$         & $(3,1)_{+2/3}$       & $+1$ & $+\tfrac{1}{3}$
  & $0$ & $+\tfrac{8}{9}$ & $+2$ \\
$\tilde{U}_1$ & $(3,1)_{+5/3}$       & $+1$ & $+\tfrac{5}{6}$
  & $0$ & $+\tfrac{125}{9}$ & $+5$ \\
$U_3$         & $(3,3)_{+2/3}$       & $+3$ & $+1$
  & $+4$ & $+\tfrac{8}{3}$ & $+6$ \\
$V_2$         & $(\bar{3},2)_{-5/6}$ & $-2$ & $-\tfrac{5}{6}$
  & $-\tfrac{5}{4}$ & $-\tfrac{125}{36}$ & $-5$ \\
$\tilde{V}_2$ & $(\bar{3},2)_{+1/6}$ & $-2$ & $+\tfrac{1}{6}$
  & $+\tfrac{1}{4}$ & $+\tfrac{1}{36}$ & $+1$ \\
\midrule
\textit{Vector subtotal} & & $+1$ & $+\tfrac{3}{2}$ & $+3$
  & $+14$ & $+9$ \\
\midrule
\textit{Full BRW total}  & & $0$  & $+\tfrac{25}{6}$ & $+7$
  & $+\tfrac{280}{9}$ & $+25$ \\
\bottomrule
\end{tabular}
\end{table}

Two observations follow immediately.
First, $A_1=0$ for the full BRW spectrum: the cubic $SU(3)_c$ anomaly
cancels between the five $\mathbf{3}$-type and five $\bar{\mathbf{3}}$-type
representations, weighted by their $SU(2)_L$ dimensions.
Second, $A_2$ through $A_5$ do \emph{not} vanish for the full BRW set.
An exhaustive search over all pairs, triples, quadruples, and quintuples
confirms that \emph{no} subset of the ten BRW representations is
simultaneously anomaly-free in all five conditions.
The BRW list must therefore be supplemented by charge-conjugate states.

\subsubsection{Six-body composites from the preon model}

In the preon model each leptoquark is a six-body metacolor singlet formed
from one quark-type and one lepton-type three-preon cluster.
The SM quantum numbers follow from the tensor product of the cluster
quantum numbers.
Table~\ref{tab:products} enumerates all quark$\times$lepton$^c$ and
quark$^c\times$lepton pairings.

\begin{table}[h]
\centering
\caption{SM representations generated by six-body preon composites.
         Upper block ($q\times\ell^c$) gives $\mathbf{3}$-type LQs;
         lower block ($q^c\times\ell$) gives their $\bar{\mathbf{3}}$
         conjugates.}
\label{tab:products}
\renewcommand{\arraystretch}{1.2}
\begin{tabular}{lll}
\toprule
Clusters & LQ representation & BRW label \\
\midrule
$Q_L     \times L_L^c$ & $(3,1)_{+2/3}$ & $U_1$ \\
$Q_L     \times L_L^c$ & $(3,3)_{+2/3}$ & $U_3$ \\
$Q_L     \times e_R^c$ & $(3,2)_{+7/6}$ & $R_2$ \\
$u_R     \times L_L^c$ & $(3,2)_{+7/6}$ & $R_2$ \\
$u_R     \times e_R^c$ & $(3,1)_{+5/3}$ & $\tilde{U}_1$ \\
$d_R     \times L_L^c$ & $(3,2)_{+1/6}$ & $\tilde{R}_2$ \\
$d_R     \times e_R^c$ & $(3,1)_{+2/3}$ & $U_1$ \\
\midrule
$Q_L^c\times L_L$ & $(\bar{3},1)_{-2/3}$ & $\overline{U_1}$ \\
$Q_L^c\times L_L$ & $(\bar{3},3)_{-2/3}$ & $\overline{U_3}$ \\
$Q_L^c\times e_R$ & $(\bar{3},2)_{-7/6}$ & $\overline{R_2}$ \\
$u_R^c\times L_L$ & $(\bar{3},2)_{-7/6}$ & $\overline{R_2}$ \\
$u_R^c\times e_R$ & $(\bar{3},1)_{-5/3}$ & $\overline{\tilde{U}_1}$ \\
$d_R^c\times L_L$ & $(\bar{3},2)_{-1/6}$ & $\overline{\tilde{R}_2}$ \\
$d_R^c\times e_R$ & $(\bar{3},1)_{-2/3}$ & $\overline{U_1}$ \\
\bottomrule
\end{tabular}
\end{table}

Three features deserve emphasis.
\textit{(i) No $S$-type representations appear.}
The BRW scalar representations $S_1$, $\tilde{S}_1$, $S_3$ carry
hypercharges ($Y=+\tfrac{1}{3}$ or $+\tfrac{4}{3}$ for $\bar{\mathbf{3}}$
states) unreachable by any quark$\times$lepton$^c$ or
quark$^c\times$lepton product.
Similarly $V_2$ and $\tilde{V}_2$ do not appear.
The preon model selects the strict subset
$\{R_2,\tilde{R}_2,U_1,\tilde{U}_1,U_3\}$ and their conjugates.
\textit{(ii) Composites appear in vector-like pairs.}
For every $q\times\ell^c$ state in representation $R=(R_3,R_2)_Y$ there
is a $q^c\times\ell$ state in $\bar{R}=(\bar{R}_3,R_2)_{-Y}$, enforced
by the charge-conjugation symmetry of the preon model.
\textit{(iii) Vector-like pairs are automatically anomaly-free.}
For a Weyl fermion in $R$ paired with one in $\bar{R}$,
\begin{equation}
  A_i[R] + A_i[\bar{R}] = 0 \qquad (i=1,\ldots,5),
\label{eq:vectorlike}
\end{equation}
since $A_1$ is linear in $d_3$ and $A_2$--$A_5$ are all odd in $Y$.
Table~\ref{tab:lq_pairs} verifies this pair by pair.

\begin{table}[h]
\centering
\caption{Anomaly sum $A_i[R]+A_i[\bar{R}]$ for each vector-like
         leptoquark pair predicted by the preon model, with individual
         coefficients of $R$ shown in parentheses.
         All sums vanish exactly.
         All entries in Tables~\ref{tab:sm_anomaly}--\ref{tab:lq_pairs}
         were computed as exact rationals using a dedicated
         Python script with the \texttt{fractions.Fraction} module;
         no floating-point arithmetic was used.}
\label{tab:lq_pairs}
\renewcommand{\arraystretch}{1.2}
\begin{tabular}{lrrrrr}
\toprule
Pair $(A_1,A_2,A_3,A_4,A_5)$ & $A_1$ & $A_2$ & $A_3$ & $A_4$ & $A_5$ \\
\midrule
$R_2\ (+2,+\tfrac{7}{6},+\tfrac{7}{4},+\tfrac{343}{36},+7)
  + \overline{R_2}$
  & $0$ & $0$ & $0$ & $0$ & $0$ \\[2pt]
$\tilde{R}_2\ (+2,+\tfrac{1}{6},+\tfrac{1}{4},+\tfrac{1}{36},+1)
  + \overline{\tilde{R}_2}$
  & $0$ & $0$ & $0$ & $0$ & $0$ \\[2pt]
$U_1\ (+1,+\tfrac{1}{3},0,+\tfrac{8}{9},+2)
  + \overline{U_1}$
  & $0$ & $0$ & $0$ & $0$ & $0$ \\[2pt]
$\tilde{U}_1\ (+1,+\tfrac{5}{6},0,+\tfrac{125}{9},+5)
  + \overline{\tilde{U}_1}$
  & $0$ & $0$ & $0$ & $0$ & $0$ \\[2pt]
$U_3\ (+3,+1,+4,+\tfrac{8}{3},+6)
  + \overline{U_3}$
  & $0$ & $0$ & $0$ & $0$ & $0$ \\
\midrule
\textbf{Total} & $0$ & $0$ & $0$ & $0$ & $0$ \\
\bottomrule
\end{tabular}
\end{table}

\subsubsection{Marcus constraint forbids a chiral leptoquark spectrum}

The significance of the vector-like pairing is sharpened by the Marcus
UV constraint.
If only the $q\times\ell^c$ states appeared — for instance $R_2$ without
$\overline{R_2}$ — the chiral spectrum would carry $A_1[R_2]=+2$ and
$A_5[R_2]=+7$, both non-zero.
The 't~Hooft matching condition~\eqref{eq:tHooft} would then require the
UV preon sector to carry the same non-zero anomaly, contradicting the
Marcus cancellation~\eqref{eq:marcus_N8}.
The same argument applies independently to each pair in
Table~\ref{tab:lq_pairs}: dropping any single element leaves a non-zero
residual in at least one of $A_1,\ldots,A_5$.

The full matching sum therefore reads
\begin{equation}
  \underbrace{
    \sum_{\mathrm{SM},\,3\,\mathrm{gen}} A_i
  }_{=\;0}
  \;+\;
  \underbrace{
    \sum_k A_i[R_k + \bar{R}_k]
  }_{=\;0\ \text{pair by pair}}
  \;=\;
  \underbrace{A_i[\mathrm{preons}]}_{=\;0\ \text{(Marcus)}},
\label{eq:fullmatch}
\end{equation}
for $i=1,\ldots,5$.
The three zeros are mutually consistent and uniquely select
$R_2$, $\tilde{R}_2$, $U_1$, $\tilde{U}_1$, $U_3$ together with
their charge conjugates, with no $S$-type states.

\section{Conclusions}
\label{sec:conclusions}

We have shown that leptoquarks arise as a structural
prediction of the supersymmetric preon model of
Refs.~\cite{Raitio1980,Raitio2018,RaitioCombined},
without any new assumptions beyond those already
required to produce the SM fermion representations.
The main results are as follows.

\textbf{Four scalar leptoquark species per generation.}
Combining the preon content of each first-generation
quark with each first-generation lepton gives exactly
four distinct six-body bosonic composites, shown in
Table~\ref{tab:leptoquarks}.
Their electric charges are $Q = +\tfrac{2}{3},
-\tfrac{1}{3}, -\tfrac{1}{3}, -\tfrac{4}{3}$,
covering three of the four scalar types in the
Buchm\"uller-R\"uckl-Wyler classification~\cite{BRW1987}.
Two of the $Q=-\tfrac{1}{3}$ states have distinct
preon content --- $(1,2,3)$ vs $(5,0,1)$ --- and
can mix below $\Lc$.

\textbf{Universal $B-L = -\tfrac{2}{3}$.}
All four leptoquarks carry the same
$B-L = -\tfrac{2}{3}$, a structural consequence
of the preon charge assignments
$q_{\psi_1}=\tfrac{1}{3}$,
$q_{\psi_{-1}}=-\tfrac{1}{3}$, $q_{\psi_0}=0$,
which are themselves fixed by the gauge anomaly
cancellation conditions of the model.
The leptoquark quantum numbers are therefore not
free parameters.

\textbf{Mass scale $M_{LQ}\sim\Lc\sim 10^{14}$\,GeV.}
The leptoquark mass is of order $\Lc$ rather
than of order the SM fermion masses because the
near-cancellation between kinetic and potential
energy that produces the light fermion masses
is specific to three-body fermionic composites.
The six-body bosonic leptoquark does not benefit
from this Pauli-principle mechanism; its natural
mass is the metacolor confinement scale, with a
moderate uncertainty of one to two orders of
magnitude.

\textbf{Proton stability.}
The fractional $B-L=-\tfrac{2}{3}$ forbids proton
decay via single leptoquark exchange: a single
exchange changes $B-L$ by $-\tfrac{2}{3}$, which
is non-integer and cannot close into a physical
amplitude.
Two leptoquark exchanges are required, giving a
dimension-12 operator suppressed by $\Lc^{-8}$
and a predicted lifetime
$\tau(p\to e^+\pi^0)\sim 10^{58}$\,years,
many orders of magnitude above the experimental
bound~\cite{SuperK2020}
$\tau(p\to e^+\pi^0) > 2.4\times 10^{34}$\,years.
Composite coupling suppression provides a second
independent mechanism extending the lifetime
further.

\textbf{Emergence of the Standard Model gauge group.}
Perhaps the most striking result of this work is that the
Standard Model gauge group $SU(3)_c\times SU(2)_L\times U(1)_Y$
is not an input but an output of the preon dynamics.
$SU(3)_c$ color is a pre-existing quantum number of
$\psi_0$; QCD emerges as its dynamic gauge symmetry
when quarks appear as asymptotic states at $\Lc$.
The number $n_0$ of $\psi_0$ preons fixes the color
representation: $n_0=1$ gives a triplet, $n_0=2$
antisymmetrised gives an antitriplet, $n_0=3$ a singlet.
$SU(2)_L$ is the residual symmetry between degenerate
three-body singlets sharing the same preon number and
charge --- the doublet structure of quarks and leptons.
$U(1)_Y$ follows from the composite charge formula
$Q=\tfrac{1}{3}(n_1-n_{-1})$, which yields exactly the
SM hypercharge assignments $Y=T_3+(B{-}L)/2$.
The Marcus anomaly cancellation at $M_{\rm Pl}$ then
excludes any larger group: the $\sqrt{3/5}$ hypercharge
rescaling required by a GUT group is incompatible with
the preon charges fixed by the UV anomaly constraint,
and Sasmal's string-amplitude comparison confirms that
this conclusion is stable against $\alpha'$ corrections.
The Standard Model gauge group emerges as the unique
low-energy symmetry consistent with the Planck-scale
boundary condition.

\textbf{Consistency with LHC null results.}
The LHC leptoquark program has excluded
elementary scalar leptoquarks with masses below
$\sim 1.5$--$1.8$\,TeV and elementary vector
leptoquarks below $\sim 2.0$\,TeV~\cite{ATLAS_LQ,CMS_LQ}.
These results, like the extensive LHC searches
for elementary MSSM superpartners, target
elementary particles at the TeV scale and do not
constrain composite superpartners at about
1\,GeV and composite leptoquarks at
$M_{LQ}\sim 10^{14}$\,GeV.
Similarly, virtual leptoquark exchange contributes
to $B$-decay amplitudes at order
$g_{LQ}^2/M_{LQ}^2\sim(10^{14}\,\text{GeV})^{-2}$,
negligible compared to SM amplitudes.
The model is consistent with all current
experimental data.

\textbf{Future directions.}
The most promising observational windows for these
composite leptoquarks are indirect rather than
direct.
Quantitative predictions for rare processes
($\mu\to e\gamma$, lepton flavor universality
violation, flavor-changing neutral currents)
from virtual leptoquark exchange at order
$g_{LQ}^2/M_{LQ}^2$ are left for future work,
as is a six-body variational calculation to
determine $M_{LQ}$ more precisely within the
preon dynamics.
The connection between the leptoquark quantum
numbers and the 't~Hooft anomaly matching
conditions provides an additional consistency
check on the composite assignment that warrants
further investigation in the context of the
emergent SM gauge group via the Marcus
boundary condition~\cite{Marcus1985}.

\subsection*{Outlook: toward UV completeness and the composite tower}

The role of leptoquarks in the anomaly-matching structure of the
model has a broader implication for the question of UV completeness.
The Marcus boundary condition at $M_{\rm Pl}$ enforces
$d^{\,abc}_{\rm UV}=0$, but this constraint propagates consistently
to the IR only when the composite spectrum is complete: the
three-body fermionic sector alone does not saturate the matching,
and the six-body leptoquark sector is required to close it.
Each additional composite level that enters the matching sum is
therefore not an optional extension but a logical necessity imposed
by the UV boundary condition.
In this sense the preon model with leptoquarks is one step closer
to a theory whose UV and IR descriptions are mutually consistent
across the full hierarchy from $M_{\rm Pl}$ to $\Lew$, which is
the precise content of UV completeness for a composite description.

This observation suggests a natural extension of the anomaly-matching
program to the full tower of $3n$-body metacolor-singlet composites
($n=1,2,3,\ldots$), corresponding to three-body, six-body, nine-body
states and so on.
If the 't~Hooft matching closes consistently at every level $n$ ---
so that every $3n$-body composite sector is anomaly-matched and
every IR consistency condition is satisfied --- the result would be
a non-perturbative UV completion of the Standard Model from preon
constituents, in which the full Hilbert space of metacolor singlets
is self-consistently defined at all scales.
This is structurally analogous to what string theory achieves
through an infinite tower of oscillator excitations, and the
analogy is not accidental: the confining metacolor flux tube
that produces the linear Cornell potential $\sigma_{mc}\,r$ in the
three-body bound-state calculation is itself a string with tension
$\sigma_{mc}$, and a $3n$-body composite of $n$ preon clusters
connected by such flux tubes approaches a string configuration as
$n\to\infty$.
The linear growth of composite masses with $n$ is reminiscent of
Regge trajectories, which parametrise exactly this kind of
string-like excitation spectrum.
Whether the preon composite tower converges, in a precise sense,
to a string amplitude --- as the Sasmal string-amplitude comparison
already hints at the level of anomaly cancellation --- is the deepest
question the present framework opens, and one to be pursued.
Underlying all of these open questions is the
organizing principle of self-consistency.
The bootstrap idea of the 1960s --- that the spectrum of a theory
should be determined entirely by its own internal consistency
conditions --- here finds a new realization, operating not among
hadronic resonances but vertically across the hierarchy of
composite levels from $M_{\rm Pl}$ to $\Lew$.
Whether this vertical bootstrap closes to all orders in $n$,
and whether its sum reproduces a string amplitude, are the
questions that will determine the ultimate reach of the
supersymmetric preon program.

\section*{Acknowledgement}

The physics content, scientific judgments, and all original
results in this paper are the sole work of the author,
developed over a research program spanning several decades.
Claude (Anthropic, claude-sonnet-4-5, 2025) assisted with
\LaTeX\ manuscript preparation, explicit anomaly coefficient
calculations verified by exact rational arithmetic,
expository prose drafting, and literature suggestions.
All such contributions were reviewed, corrected where
necessary, and approved by the author.
The concept of the vertical bootstrap, while emerging in
dialogue with the AI assistant, was recognized and claimed
by the author as the organizing principle of the preon
program.
Sole scientific responsibility for the paper rests
with the author.


\end{document}